\documentclass[preprint,showpacs,showkeys,preprintnumbers,amsmath,amssymb]{revtex4}
\usepackage{amssymb}
\usepackage[dvips]{graphicx}

\begin{document}

\title{B\"{a}cklund transformations for Burgers Equation via localization of residual symmetries }

\author{ Xi-zhong Liu$^{a}$, Jun Yu$^{a}$, Bo Ren$^{a}$}

\affiliation{$^{a}$Institute of Nonlinear Science, Shaoxing University, Shaoxing 312000, China
}

\begin{abstract}
In this paper, we obtained the non-local residual symmetry related to truncated Painlev\'{e} expansion of Burgers equation. In order to localize the residual symmetry, we introduced new variables to prolong the original Burgers equation into a new system. By using Lie's first theorem, we got the finite transformation for the localized residual symmetry. More importantly, we also localized the linear superposition of multiple residual symmetries to find the corresponding finite transformations. It is interesting to find that the $n^{th}$ B\"{a}cklund transformation for Burgers equation can be expressed by determinants in a compact way.
\end{abstract}

\pacs{02.30.Jr,\ 02.30.Ik,\ 05.45.Yv,\ 47.35.Fg}

\keywords{Burgers equation, residual symmetry, B\"{a}cklund transformation}

\maketitle
\section{Introduction}
The integrable equation has wide applications in the field of nonlinear science, like nonlinear optics, the theory of deep water waves and plasma physics, etc. It is well known that there are many ways to obtain explicit solutions of integrable equations, such as the Inverse Scattering transformation, the Hirota technique, symmetry reduction, B\"{a}cklund transformation (BT), Darboux
transformation (DT), the algebra-geometric method \cite{matv,hiro,abl,wada,roger,guch}. Among which, the BT and DT is the most direct and efficient approach for the construction of exact solutions (see, e.g., \cite{darboux,levi}). It is interesting that, through iterations, one is often led to compact representations in terms of special determinants such as the Wronskian or Grammian for N-soliton solutions. The essential idea of BT or DT is to construct the solution of integrable equations by the seed solutions, which means that there exist underlying symmetries related to these transformations for the equations under consideration.

One of the most powerful method to prove the integrability of a model is the Painlev\'{e} analysis developed
by WTC (Weiss-Tabor-Canvela). For many Painlev\'{e} integrable equation, one can get corresponding  truncated Painlev\'{e} expansion which serving as as a BT for the original equation. It is known that to search for non-local symmetries is of considerable interest. Recent studies \cite{huxiao,louresi} find that the symmetries related to  truncated Painlev\'{e}expansion is just the residue of the expansion with respect to singular manifold, so it is called residual symmetry. Like other symmetries related to BT, the residual symmetry is nonlocal and many research has been done to construct new solutions by using these symmetries. One of the direct way is to prolong the original equation to localize the nonlocal symmetries and then obtain new symmetry reduction solutions by standard Lie symmetry approach \cite{loukp}. In this paper, we will get BT for Burgers equation by localizing the residual symmetry and then generalize the procedure to find the $n^{th}$ BT.

The Burgers equation
\begin{equation}\label{bur}
u_t+uu_x=\nu u_{xx}.
\end{equation}
with viscosity coefficient $\nu$, is one of the most simplest integrable equation. It serves as a fundamental equation in the field of fluid mechanics and also occurs in various areas of applied physics such as the modeling of gas dynamics and traffic flow. The exact solutions of Burgers equation have been studied by many authors. In Ref. \cite{lou4}, the infinite many symmetries and exact solutions of Burgers equation are obtained by the repeated symmetry reduction approach.

The paper is organized as followings. In Sec. II, we get the non-local residual symmetry of Burgers equation and then localize it in a new prolonged system. The corresponding finite transformation group is obtained by using Lie's first theorem, consequently. In Sec. III, we first localize the linear superposition of multiple residual symmetries by introducing various new variables, then we get the $n^{th}$ BT of the prolonged Burgers system formulated in terms of determinants.

\section{localization of residual symmetry and related B\"{a}cklund transformation}

As the Burgers Eq. \eqref{bur} being Painlev\'{e} integrable, we search for the truncated Painlev\'{e} expansion in the general form
\begin{equation}\label{genpain}
u=\sum_{i=0}^{\alpha}{u_i\phi^{i-\alpha}},
\end{equation}
where $u_\alpha$ is an arbitrary solution of the discussed equation while $u_{\alpha-1},u_{\alpha-2},\cdots, u_0$ are all related to derivatives of $\phi$. After substituting $u\sim u_0\phi^\alpha$ into Eq. \eqref{bur} and balancing the nonlinear and dispersion terms, we have the truncated Painlev\'{e} expansion as
\begin{equation}\label{pain1}
u=\frac{u_0}{f}+u_1
\end{equation}
with $u_0$ and $u_1$ to be determined.

Now substituting Eq. \eqref{pain1} into Eq. \eqref{bur}, we get
\begin{multline}\label{subspain1}
u_{1t}+u_1u_{1x}-\nu u_{1xx}+\phi^{-1}(u_0u_{1x}+u_1u_{0x}+u_{0t}-\nu u_{0xx})+\phi^{-2}(-u_0\phi_t+2\nu u_{0x}\phi_x+\nu u_0\phi_{xx}\\+u_0u_{0x}-u_1u_0\phi_x)-\phi^{-3}\phi_xu_0(2\nu\phi_x+u_0)=0,
\end{multline}
Vanishing the coefficients of $\phi^{-3}$ and $\phi^{-2}$, we get
\begin{equation}\label{u0}
u_0 = -2\nu \phi_x,
\end{equation}
and
\begin{equation}\label{u1}
u_1 = \frac{-\phi_t+\nu\phi_{xx}}{\phi_x}.
\end{equation}
While from the coefficients of $\phi^0$ and $\phi^{-1}$, we see that $u_0$ is a symmetry of Burgers equation \eqref{bur} with $u_1$ as a solution.

Substituting Eq. \eqref{u1} into the coefficient of $\phi^0$ in Eq. \eqref{subspain1}, we get the Schwarz form of Burgers equation as
\begin{equation}\label{sch}
K_t-KK_x-2\nu K_{xx}+\nu^2\{\phi; x\}_x=0,
\end{equation}
where the Schwarzian derivative $\{\phi; x\}=\frac{\phi_{xxx}}{\phi_x}-\frac{3\phi_{xx}^2}{2\phi_x^2}$ and $K=\frac{\phi_t}{\phi_x}$ are all invariant under Mobious transformation, i.e.
\begin{equation}
\phi\to \frac{a_1\phi+b_1}{a_2\phi+b_2},\, a_1a_2 \neq b_1b_2.
\end{equation}
In other words, Eq. \eqref{sch} possesses three symmetries $\sigma^{\phi}=d_1$, $\sigma^{\phi}=d_2\phi$ and
$$\sigma^{\phi}=d_3\phi^2$$
with arbitrary constants $d_1,\,d_2$ and $d_3.$

Since $u_0$ is a non-local symmetry relating to the BT \eqref{pain1}, one can naturally believe that it can be localized to a Lie point symmetry such that we can use the Lie' s first theorem to recover the original BT. To this end, we prolong the Burgers equation into the following system
\begin{subequations}\label{enlar}
\begin{equation}
u_t+uu_x-\nu u_{xx}=0,
\end{equation}
\begin{equation}
u= \frac{-\phi_t+\nu\phi_{xx}}{\phi_x},
\end{equation}
\begin{equation}\label{gh1}
g= \phi_x,
\end{equation}
\end{subequations}
and get the localized Lie point symmetry
\begin{subequations}\label{pointsy1}
\begin{equation}
\sigma^{u} = g,
\end{equation}
\begin{equation}
\sigma^g = \frac{\phi g}{\nu},
\end{equation}
\begin{equation}
\sigma^{\phi} = \frac{\phi^2}{2\nu},
\end{equation}
\end{subequations}
where $d_3$ is fixed as $\frac{1}{2\nu}$. It can be easily verified that Eq. \eqref{pointsy1} indeed satisfies the linearized form of \eqref{enlar}
\begin{subequations}\label{linear}
\begin{equation}
\sigma^{\phi}_{x}-\sigma^{g}=0,
\end{equation}
\begin{equation}
\sigma^{u}_t+\sigma^{u}u_x+u\sigma^{u}_x-\nu\sigma^{u}_{xx}=0,
 \end{equation}
\begin{equation}
\sigma^{\phi}_x\phi_t-\sigma^u\phi_x^2+(\nu\sigma^{\phi}_{xx}-\sigma^{\phi}_t)\phi_x-\nu\sigma^{\phi}_x \phi_{xx}=0.
\end{equation}
\end{subequations}

The result \eqref{pointsy1} indicates that the residual symmetries \eqref{u0} is localized in
the properly prolonged system \eqref{enlar} with the Lie point symmetry vector
\begin{equation}\label{pointV}
V=g\partial_{u}+\frac{\phi g}{\nu}\partial_{g}+\frac{\phi^2}{2\nu}\partial_{\phi}.
\end{equation}

Now we can get the finite transformation of Lie point symmetry \eqref{pointV} straightforwardly, which can be stated in the following theorem

\noindent\emph{ \textbf{Theorem 1.}}
If $\{u,g,\phi\}$ is a solution of the prolonged system \eqref{enlar}, then so is $\{\hat{u},\hat{g},\hat{\phi}\}$ with
\begin{subequations}\label{pointsy}
\begin{equation}
\hat{u}=u-\frac{2\epsilon\nu g}{\epsilon\phi-2\nu},
\end{equation}
\begin{equation}
\hat{g} =\frac{4\nu^2g}{(\epsilon\phi-2\nu)^2},
\end{equation}
\begin{equation}
\hat{\phi} = -\frac{2\nu\phi}{\epsilon\phi-2\nu},
\end{equation}
\end{subequations}
where $\epsilon$ is an arbitrary group parameter.

\emph{Proof.} Using Lie's first theorem on vector \eqref{pointV} with the corresponding
 initial condition as follows
\begin{eqnarray}
\\ \frac{{\rm d}\hat{u}(\epsilon)}{{\rm d}\epsilon}&=& \hat{g}(\epsilon),\,\quad \hat{u}(0)=u,\\
\frac{{\rm d}\hat{g}(\epsilon)}{{\rm d}\epsilon}&=&\frac{\hat{\phi}(\epsilon)\hat{g}(\epsilon)}{\nu},\,\quad \hat{g}(0)=g,\\
\frac{{\rm d}\hat{\phi}(\epsilon)}{{\rm d}\epsilon}&=&\frac{\hat{\phi}^2(\epsilon)}{2\nu},\,\quad \hat{\phi}(0)=\phi,
\end{eqnarray}
one can easily obtain the solutions of the above equations stated in Theorem 1, thus the theorem is
proved.
\section{B\"{a}cklund transformation related to multiple residual symmetries}
From the expression \eqref{u0}, we known that the residual symmetry depends only on the solution of original equation in Schwarz form, i.e. Eq. \eqref{sch} for Burgers equation. So there exist infinitely many residual symmetries expressed by different solutions of Schwarzian equation and any linear superposition of these symmetries is also a symmetry of the original equation. As for Burgers equation, the general form of a residual symmetry takes the form
\begin{equation}\label{gensym}
\sigma_u^n=\sum_{i=1}^{n}c_i\phi_{i,x},
\end{equation}
for arbitrary positive integer $n$, where $\phi_i, i=1,\cdots, n$ are different solutions of Eq. \eqref{sch}.

We can also localize the residual symmetry \eqref{gensym} by prolonging the Burgers system \eqref{enlar} further, which can be stated in the following theorem:

\noindent\emph{ \textbf{Theorem 2.}}  The symmetry \eqref{gensym} is localized to a Lie point symmetry
\begin{subequations}\label{locmulti}
\begin{equation}
\sigma^u_n=\sum_{i=1}^{n}c_ig_{i},
\end{equation}
\begin{equation}
\sigma^{\phi_i}=\frac{c_i}{2\nu}\phi_i^2+\frac{1}{2\nu}\sum_{j\neq i}^nc_j\phi_j\phi_i,
\end{equation}
\begin{equation}
\sigma^{g_i}=\frac{c_i}{\nu}g_i\phi_i+\frac{1}{2\nu}\sum_{j\neq i}^nc_j(g_i\phi_j+g_j\phi_i),
\end{equation}
\end{subequations}
where $\{u,\,\phi_i,\,g_i\}$ is a solution of the enlarged system
\begin{subequations}\label{prolongbur}
\begin{equation}
u_t+uu_x-\nu u_{xx}=0,
\end{equation}
\begin{equation}\label{uphii}
u = \frac{-\phi_{i,t}+\nu\phi_{i,xx}}{\phi_{i,x}},
\end{equation}
\begin{equation}
g_i=\phi_{i,x}.
\end{equation}
\end{subequations}
\emph{Proof.} It is easy to write down the linearized form of the enlarged system \eqref{prolongbur} 
\begin{subequations}
\begin{equation}
\sigma^{u}_t+\sigma^{u}u_x+u\sigma^{u}_x-\nu\sigma^{u}_{xx}=0,
 \end{equation}
 \begin{equation}\label{sigmau}
\sigma^u = \frac{\nu\sigma^{\phi_i}_{xx}-\sigma^{\phi_i}_t}{\phi_{i,x}}-\frac{(\nu\phi_{i,xx}-\phi_{i,t})
\sigma^{\phi_i}_x}{\phi_{i,x}^{2}}
\end{equation}
\begin{equation}\label{sigmagi}
\sigma^{g_i}=\sigma^{\phi_i}_x, \, i=1,\cdots, n.
\end{equation}
\end{subequations}
To prove this theorem, we first consider the special case, i.e., for any fixed $k$, $c_k\neq 0$ while $c_j=0, j\neq k$ in Eq. \eqref{gensym}. In this case, we get the localized symmetry for $u,\ \phi_k$ and $g_k$ from Eq. \eqref{pointsy1}
\begin{subequations}\label{pointsyn}
\begin{equation}\label{sigmaug}
\sigma^{u} = c_kg_k,
\end{equation}
\begin{equation}
\sigma^{\phi_k} = \frac{c_k\phi_k^2}{2\nu},
\end{equation}
\begin{equation}
\sigma^{g_k} = \frac{c_k\phi_k g_k}{\nu},
\end{equation}
\end{subequations}
For $j\neq k$, we eliminate $u$ through Eq. \eqref{uphii} by taking $i=k$ and $i=j$, respectively. Then we have
\begin{equation}\label{phixx}
\phi_{j,xx} = -\frac{\phi_{j,x}\phi_{k,t}}{\nu\phi_{k,x}}+\frac{\phi_{j,x}\phi_{k,xx}}{\phi_{k,x}}+\frac{\phi_{j,t}}
{\nu}.
\end{equation}
Substituting Eq. \eqref{sigmaug} into Eq. \eqref{sigmau} with $i=j$ and vanishing $\phi_{j,xx}$ through \eqref{phixx}, we get
\begin{equation}
\sigma^{\phi_j}=\frac{1}{2\nu}c_k\phi_k\phi_j,
\end{equation}
then the symmetry for $g_j$ can be easily obtained from \eqref{sigmagi} with $i=j$
\begin{equation}
\sigma^{g_j}=\frac{1}{2\nu}c_k(g_j\phi_k+g_k\phi_j).
\end{equation}
From the linear superposition property of symmetry, we get the desired result \eqref{locmulti} by taking linear combination of the above results for $k=1,\cdots, n$, the theorem is thus proved.

Now, we can obtain the finite transformations according to the localized symmetry \eqref{locmulti} by using the Lie's first theorem and the corresponding initial value problem takes the form
\begin{subequations}\label{inival}
\begin{equation}
\frac{{\rm d}U(\epsilon)}{{\rm d}\epsilon}=\sum_{j=1}^{n}c_j\Phi_{j,x}(\epsilon),
\end{equation}
\begin{equation}
\frac{{\rm d}\Phi_i(\epsilon)}{{\rm d}\epsilon}=\frac{c_i}{2\nu}\Phi_i^2(\epsilon)+\frac{1}{2\nu}\sum_{j\neq i}^nc_j\Phi_j(\epsilon)\Phi_i(\epsilon)
\end{equation}
\begin{equation}
\frac{{\rm d}G_i(\epsilon)}{{\rm d}\epsilon}=\frac{c_i}{\nu}G_i(\epsilon)\Phi_i(\epsilon)+\frac{1}{2\nu}\sum_{j\neq i}^nc_j(G_i(\epsilon)\Phi_j(\epsilon)+G_j(\epsilon)\Phi_i(\epsilon))
\end{equation}
\begin{equation}
U(0)=u,\, \Phi_i(0)=\phi_i,\, G_i(0)=g_i,\ i=1,\cdots ,n.
\end{equation}
\end{subequations}
By solving out \eqref{inival}, we get the following $n^{th}$ B\"{a}cklund transformation theorem.

\noindent\emph{ \textbf{Theorem 3.}} If $\{u,\ \phi_i,\ g_i,\ i=1,\ \cdots, n\}$ is a solution of the prolonged Burgers system \eqref{prolongbur}, so is $\{U(\epsilon),\ \Phi_i(\epsilon),\ G_i(\epsilon),\ i=1,\ \cdots, n\}$ where
\begin{subequations}
\begin{equation}
U(\epsilon)=u-2\nu(\ln\Delta)_x,
\end{equation}
\begin{equation}
\Phi_i(\epsilon)=-2\nu\frac{\Delta_i}{\Delta},
\end{equation}
\begin{equation}
G_i(\epsilon)=\Phi_{i,x}(\epsilon)
\end{equation}
\end{subequations}
with $\Delta$ and $\Delta_i$ are the determinants of
\begin{equation}
\left(
\begin{array}{cccccc}
\frac{c_1}{2\nu}\epsilon\phi_1-1 & c_1\epsilon \mu_{1,2} & \cdots&c_1\epsilon \mu_{1,j} & \cdots & c_1\epsilon  \mu_{1,n}\\
c_2\epsilon\mu_{1,2} & \frac{c_2}{2\nu}\epsilon\phi_2-1 & \cdots&c_2\epsilon \mu_{2,j} & \cdots & c_2\epsilon  \mu_{2,n}\\
\vdots &\vdots&\vdots&\vdots&\vdots&\vdots\\
c_j\epsilon\mu_{1,j} &c_j\epsilon\mu_{2,j} &\cdots& \frac{c_j}{2\nu}\epsilon\phi_j-1 & \cdots & c_j\epsilon\mu_{j,n}\\
\vdots &\vdots&\vdots&\vdots&\vdots&\vdots\\
c_n\epsilon\mu_{1,n}&c_n\epsilon  \mu_{2,n}&\cdots&c_n\epsilon\mu_{j,n}&\cdots&\frac{c_n}{2\nu}\epsilon\phi_n-1
\end{array}
\right),\,\ \mu_{i,j}=\frac{1}{2\nu}\sqrt{\phi_i\phi_j}
\end{equation}
and
\begin{equation}
\left(
\begin{array}{cccccccc}
\frac{c_1}{2\nu}\epsilon\phi_1-1 & c_1\epsilon \mu_{1,2} & \cdots&c_1\epsilon \mu_{1,i-1} &c_1\epsilon \mu_{1,i} &c_1\epsilon \mu_{1,i+1} & \cdots & c_1\epsilon  \mu_{1,n}\\
c_2\epsilon\mu_{1,2} & \frac{c_2}{2\nu}\epsilon\phi_2-1 & \cdots&c_2\epsilon \mu_{2,i-1} &c_2\epsilon \mu_{2,i} &c_2\epsilon \mu_{2,i+1} & \cdots & c_2\epsilon  \mu_{2,n}\\
\vdots &\vdots&\vdots&\vdots&\vdots&\vdots&\vdots&\vdots\\
c_{i-1}\epsilon\mu_{1,i-1} &c_{i-1}\epsilon\mu_{2,i-1} &\cdots& \frac{c_{i-1}}{2\nu}\epsilon\phi_{i-1}-1&c_{i-1}\epsilon\mu_{i-1,i}&c_{i-1}\epsilon\mu_{i-1,i+1} & \cdots & c_{i-1}\epsilon\mu_{i-1,n}\\
\mu_{1,i}&\mu_{2,i}&\cdots&\mu_{i-1,i}&\frac{1}{2\nu}\phi_i&\mu_{i,i+1}&\cdots&\mu_{i,n}\\
c_{i+1}\epsilon\mu_{1,i+1} &c_{i+1}\epsilon\mu_{2,i+1} &\cdots& c_{i+1}\epsilon\mu_{i-1,i+1}-1&c_{i+1}\epsilon\mu_{i,i+1}&\frac{c_{i+1}}{2\nu}\epsilon\phi_{i+1}-1 & \cdots & c_{i+1}\epsilon\mu_{i+1,n}\\
\vdots &\vdots&\vdots&\vdots&\vdots&\vdots\\
c_n\epsilon\mu_{1,n}&c_n\epsilon  \mu_{2,n}&\cdots&c_n\epsilon\mu_{i-1,n}&c_n\epsilon\mu_{i,n}&c_n\epsilon\mu_{i+1,n}&\cdots&\frac{c_n}{2\nu}\epsilon\phi_n-1
\end{array}
\right).
\end{equation}

From theorem 3, we can get infinitely many new solutions from any any seed solutions of Burgers equation and its Schwarz form. For example, taking the following kink soliton as seed solution for Burgers equation \cite{jinyan}, i.e. 
\begin{equation}\label{uspecial}
u=-2k\nu[1+\tanh(kx+2\nu k^2t)]
\end{equation}
with an arbitrary constant $k$, we get the solution of Schwarzian Burgers equation \eqref{sch} by solving out Eq. \eqref{uphii}, which takes the form
\begin{equation}\label{specialphi}
\phi_i = A_i\tanh(kx+2\nu k^2t)
\end{equation}
with arbitrary constants $A_i$, and then
\begin{equation}\label{muij}
\mu_{i,j}=\frac{\sqrt{A_iA_j}}{2\nu}\tanh(kx+2\nu k^2t).
\end{equation}
Substituting Eqs. \eqref{uspecial}, \eqref{specialphi} and \eqref{muij} into theorem 3, we get a new kink soliton solution for Burgers equation
\begin{equation}
u=-\frac{2k\nu[(c_1\epsilon A_1+c_2\epsilon A_2-2\nu)\tanh(kx+2\nu k^2t)+c_1\epsilon A_1+c_2\epsilon A_2-2\nu]}{\epsilon(c_1A_1+c_2A_2)\tanh(kx+2\nu k^2t)-2\nu}
\end{equation}
for $n=2$ case.

\section{Conclusion and discussion}
To sum up, the non-local residual symmetry of Burgers equation related to truncated Painlev\'{e} expansion can be localized if we introduce a new variable $g$ to eliminate the derivative of $\phi$. Using the Lie's first theorem to the Lie point symmetry of the prolonged system $\{u, \phi, g\}$, one can obtain the generalized BT for Burgers system which include the truncated Painlev\'{e} expansion as its special one.

Furthermore, we also obtained the finite transformations for linear superposition of multiple residual symmetries, which is equivalent to the $n^{th}$ BT for Burgers. From theorem 3, It is interesting to find that the $n^{th}$ BT can be constructed in terms of determinants in a compact way and various solutions, including soliton solutions, can be obtained through it if the seed solutions are given. It should be noted that the commutable property for BT or DT becomes trivial in the framework of symmetry, because any symmetry group allows commutativity law in addition algorithm.

In this paper, we discussed the only for the localization of residual symmetry for constructing new BT for Burgers equation. It is obvious that the the same procedure can be executed for other nonlocal symmetries, such as those obtained from B\"{a}cklund transformation, the bilinear forms and negative hierarchies, the nonlinearizations \cite{cao,cheng} and self-consistent sources \cite{zeng} etc. We believe that the idea is rather general and could be applied to other physical interested models as well.

\begin{acknowledgments}
This work was supported by the National Natural Science Foundation of China under Grant Nos. 11347183,10875078 and 11305106,, the Natural Science Foundation of Zhejiang Province of China Grant Nos. Y7080455 and LQ13A050001.
\end{acknowledgments}

\end{document}